\documentstyle[preprint,aps,epsf,eqsecnum]{revtex}
\begin{document}
\tighten

\newcommand{\be}{\begin{equation}}
\newcommand{\ee}{\end{equation}}
\newcommand{\bea}{\begin{eqnarray}}
\newcommand{\eea}{\end{eqnarray}}
\newcommand{\r}{\rangle }
\newcommand{\la}{\langle }
\draft

%\preprint{winnipeg }

\title {Nonlinear Response from Transport Theory and Quantum Field Theory at
Finite Temperature}

\author{ M.E. Carrington${}^{a,b}$, Hou Defu${}^{a,b,c}$ and R.
Kobes${}^{b,d}$}

 \address{ ${}^a$ Department of Physics, Brandon University, Brandon, 
Manitoba,
R7A 6A9 Canada\\
 ${}^b$  Winnipeg Institute for Theoretical Physics, Winnipeg, Manitoba \\
${}^c$ Institute of Particle Physics, Huazhong Normal University, 430070 Wuhan,
China \\
${}^d$ University of Winnipeg, Winnipeg, Manitoba, R3B 2E9 Canada }

%\author{}
%\address{}

\date{\today}
\maketitle

\begin{abstract}

We study nonlinear response in weakly coupled hot $\phi^4$ theory.  We obtain
an expression for a quadratic shear viscous response coefficient using two
different formalisms: transport theory and response theory. The transport
theory calculation is done by assuming a local equilibrium form for the
distribution function and expanding in the gradient of the local four
dimensional velocity field.  By doing a gradient expansion on the Boltzmann
equation we obtain a hierarchy of equations for the coefficients of this
expansion.

To do the response theory calculation we use
Zubrave's techniques in nonequilibrium statistical mechanics to derive a
generalized Kubo formula. Using this formula allows us to obtain the quadratic
shear viscous response from the three-point retarded green function of the
viscous shear stress tensor.   We use the closed time path formalism of real
time finite temperature field theory to show
that this three-point function can be calculated  by writing it as an integral equation involving a four-point
vertex.  This four-point vertex can in turn be obtained from an integral
equation which represents the resummation of an infinite series of ladder and
extended-ladder diagrams.

The connection between transport theory and response theory is made when we
show that the integral equation for this four-point vertex has exactly the same
form as the equation obtained from the Boltzmann equation for the 
coefficient of
the quadratic term of the gradient expansion of the distribution function.  We
conclude that calculating the quadratic shear viscous response using transport
theory and keeping terms that are quadratic in the gradient of the velocity
field in the expansion of the Boltzmann equation is equivalent to calculating
the quadratic shear viscous response from response theory using the
next-to-linear response Kubo formula, with a vertex given by an infinite
resummation of ladder and extended-ladder diagrams.

\end{abstract}
% \pacs{PACS numbers: 11.10Wx, 11.15Tk, 11.55Fv}
% \maketitle

%\narrowtext
%%%%%%%%%%%%%%%%%%%%%%%%%%%%%%%%%%%%%%%%%%%%%%%%%%%%%%%%%%%%%%%%%%

\section{Introduction}

    Fluctuations occur in a system perturbed slightly away from equilibrium.
The responses to these fluctuations are described by transport coefficients
which characterize the dynamics of long wavelength, low frequency 
fluctuations in the medium \cite{eh,Groot}.
The investigation of transport coefficients in high temperature
gauge theories is important in cosmological applications
such as electroweak baryogenesis \cite{bg} and in the context of heavy ion
collisions \cite{hvi}.

 There are two basic methods to calculate transport coefficients: transport
theory and  response theory
\cite{gyu,kubo,MartinK,Hosoya,Horsley,jeon1,hu,llog}.  Using the transport
theory method one starts from a local equilibrium form for the distribution
function and performs an expansion in the gradient of the four-velocity field.
The coefficients of this expansion are determined from the classical Boltzmann
equation \cite{jeon1,llog}.
In the response theory approach one divides the Hamiltonian into a bare piece
and a perturbative piece that is linear in the gradient of the four-velocity
field.  One uses standard perturbation theory to obtain the Kubo formula for
the viscosity in terms of retarded green functions \cite{kubo}.  These green
functions are then evaluated using equilibrium quantum field theory.  As is
typical in finite temperature field theory, it is not sufficient to calculate
perturbatively in the coupling constant: there are certain infinite sets of
diagrams that contribute at the same order in perturbation theory and have to
be resummed \cite{pisarski,LAD}.

In this paper, we want to compare these two methods.  The response theory
approach allows us to calculate transport coefficients from first principles
using the well understood methods of quantum field theory. On the other hand,
the transport theory approach involves the use of the Boltzmann equation which
is itself derived from some more fundamental theory using, among other things,
the quasiparticle approximation.  In this sense, the response theory approach
is more fundamental than the transport theory method. However, the response
theory approach can be quite difficult to implement, even for a high
temperature weakly coupled scalar theory, because of the need to resum infinite
sets of diagrams \cite{pisarski,LAD}.  These considerations motivate us to
understand more precisely the connection between the more practical transport
theory method, and the more fundamental response theory approach.

Some progress has already been made in this direction.  It has been shown that
keeping only terms which are linear in the gradient expansion in the transport
theory calculation is equivalent to using the linear response approximation to
obtain the usual Kubo formula for the shear viscosity in terms of a retarded
two-point function, and calculating that two-point function using standard
equilibrium quantum field theory techniques to resum an infinite set of ladder
diagrams \cite{jeon1,meg,enke}.  This result is not surprising since it has
been known for some time that ladder diagrams give large contributions to
$n$-point functions with ultra soft external lines \cite{LAD}.

To date, calculations of transport coefficients have been limited to linear
response. In some physical situations however nonlinear response can be
important, especially for relativistic  gauge
theories \cite{bdc,ASY,lm,blz}. It is therefore of interest to study nonlinear
response.
In this paper we study nonlinear response to fluctuations in weakly coupled
high temperature scalar $\phi^4$ theory using both the transport method and the
response theory approach.  We study the relationship between these two
approaches at the level of quadratic response.

This paper will be organized as follows. In section II we define shear
viscosity and quadratic shear viscous response using a hydrodynamic expansion
of the energy-momentum tensor which includes up to quadratic terms in the
gradient of the four-velocity.
In section III we calculate viscosity using the transport theory method by
performing a gradient expansion on the Boltzmann equation and obtaining a
hierarchy of equations.  In section IV we derive the quadratic response Kubo
formula by using
Zubrave's techniques in nonequilibrium statistical mechanics.  We obtain an
expression which relates the quadratic shear viscous response coefficient  to
the three-point retarded green function of the viscous shear stress tensor.
Starting from this generalized Kubo formula we calculate the quadratic shear
viscous response using standard techniques of finite temperature quantum field
theory. We show
that the quadratic shear viscous response can be obtained as an integral over a
four-point vertex. This four-point vertex  satisfies an integral equation
involving terms which are quadratic in a retarded three-point  function, which
itself
satisfies a linear integral equation. These
two integral equations represent the resummation of an infinite series of
ladder and
extended-ladder diagrams. We show that
these integral equations, which represent ladder and extended-ladder
resummations,
 have  the same form as the first two equations in the hierarchy obtained from
expanding the Boltzmann equation.
We discuss our results and present our conclusions section V.

\section{Viscosity}

In a system that is out of equilibrium, the existence of gradients
in thermodynamic parameters like the temperature and the four dimensional
velocity field give rise to thermodynamic forces. These thermodynamic forces
lead to  deviations from the equilibrium expectation value of the energy
momentum tensor which are characterized by transport coefficients like the
thermal conductivity and the shear and bulk viscosities.   In order to separate
these different physical processes we decompose the energy-momentum tensor as,
\be
T^{\mu\nu}=\epsilon u^\mu u^\nu-p\Delta^{\mu\nu}+P^\mu u^\nu +P^\nu u^\mu
+\pi^{\mu\nu}\,;~~~~\Delta_{\mu\nu}=g_{\mu\nu}-u_\mu u_\nu.
\ee
The quantities $\epsilon$, $p$, $P_\mu$ and $\pi_{\mu\nu}$ have the physical
meanings of internal energy density, pressure, heat current and viscous shear
stress, respectively.  The four vector $u_\mu(x)$ is the four dimensional
four-velocity field which satisfies $u^\mu(x) u_\mu(x)=1$.  The expansion
coefficients are given by
\bea
\epsilon&=&u_\alpha u_\beta T^{\alpha\beta}\nonumber\\
p&=&-\frac{1}{3}\Delta_{\alpha\beta}T^{\alpha\beta}\nonumber\\
P_\mu&=&\Delta _{\mu\alpha}u_{\beta} T^{\alpha\beta} \label{def1}\\
\pi_{\mu\nu}&=&(\Delta_{\mu\alpha}\Delta_{\nu\beta}-
\frac{1}{3}\Delta_{\nu\mu}\Delta_{\alpha\beta})T^{\alpha\beta}.\nonumber
\eea

The viscosity is obtained from the expectation value of the viscous shear
stress part of the energy momentum tensor.  We expand in gradients of the
four-velocity field and write,
\bea
&& \delta \langle  \pi_{\mu\nu}\rangle =\eta^{(1)} H_{\mu\nu} + \eta^{(2)}
H^{T2}_{\mu\nu} + \cdots  \label{DEF} \\
&& H_{\mu\nu} = \partial_\mu u_\nu + \partial_\nu u_\nu - \frac{2}{3}
\Delta_{\mu\nu} \Delta_{\rho\sigma}
\partial^\rho  u^\sigma \nonumber \\
&&
H_{\mu\nu}^{T2}:=H_{\mu\rho}H^\rho_{~\nu}-
\frac{1}{3}\Delta_{\mu\nu}H_{\rho\sigma}
H^{\rho\sigma} \nonumber
\eea
where
$\eta^{(1)}$ and $\eta^{(2)}$ are the coefficients of the terms that are linear
and quadratic respectively in the gradient of the four-velocity. The first
coefficient  is the usual shear viscosity.  The second has has not been widely discussed in the literature -- we shall call it the quadratic shear 
viscous response.

Throughout this paper we work with $\phi^4$ theory.  The Lagrangian for this
theory is
\bea
{\cal L} = \frac{1}{2}[(\partial_\mu \phi)^2 - m^2 \phi^2] -
\frac{\lambda}{4!}\phi^4
\eea
and the coupling constant is assumed to be small: $\lambda \ll 1$.

\section{Viscosity from Transport Theory}

Kinetic theory and the Boltzmann equation can be used to calculate transport
properties of dilute many-body systems.  One assumes that, except during brief
collisions, the system can be considered as being composed of classical
particles with well defined position, energy and momentum.  This picture is
valid when the mean free path is large compared with the Compton wavelength of
the particles.  At high temperature the typical mean free path of thermal
excitations is ${\cal O}(1/\lambda^2T)$ and is always larger than the typical
Compton wavelength of effective thermal oscillations which is 
${\cal O}(1/\sqrt{\lambda}T)$ \cite{jeon1}.
 We introduce a phase space distribution function
$f(x,\underline{k})$ which describes the evolution of the phase
space probability density for the fundamental particles comprising a fluid. In
this expression and in the following equations the underlined momenta  are on
shell, since we are describing a system of particles.  The form for
$f(x,\underline{k})$ in local equilibrium is,
\be
f^{(0)}=\frac{1}{e^{\beta(x) u_\mu(x) \underline{k}^\mu}-1}:=n_k\,;~~N_k:=1+2n_k
\,.\label{fequib}
\ee
We study the Boltzmann equation in the hydrodynamic regime where we consider
times which are long compared to the mean free time and describe the relaxation
of the system in terms of long wavelength fluctuations in locally conserved
quantities.  For a simple fluid without any additional conserved charges, the
only locally conserved quantities are energy and momentum.
To solve the Boltzmann equation in this near equilibrium hydrodynamic regime, 
we expand the distribution function around the local equilibrium form using a
gradient expansion.  We go to a local rest frame in which we can write
$\vec{u}(x)=0$.  Note that this does not imply that gradients of the form
$\partial_i u_j$ must be zero.  In the local rest frame (\ref{DEF}) becomes,
\bea
&& \delta \langle  \pi_{ij}\rangle = -\eta^{(1)} H_{ij} + \eta^{(2)}
H^{T2}_{ij} + \cdots  \label{DEFB} \\
&& H_{ij} = \partial_i u_j + \partial_j u_i - \frac{2}{3} \delta_{ij}
(\vec{\partial} \cdot \vec{u}) \nonumber \\
&& H_{ij}^{T2}:=H_{ik}H^k_{~j}-\frac{1}{3}\delta_{ij}~H_{lm}
H_{lm} \nonumber
\eea
In all of the following expressions we keep only linear terms that contain one
power of $H_{ij}$ and quadratic terms that contain two powers of $H_{ij}$,
since these are the only terms that contribute to the viscosity coefficients we
are trying to calculate.

We write,
\bea
f=f^{(0)} + f^{(1)} + f^{(2)} + \cdots
\label{expf}
\eea
with,
\bea
f^{(1)} \sim \underline{k}_\mu \partial^\mu f^{(0)}\,;~~~~f^{(2)} \sim
\underline{k}_\mu \partial^\mu f^{(1)}.
\eea
Using (\ref{fequib}) we obtain,
\bea
&&\underline{k}_\mu \partial^\mu f^{(0)} = -\beta\, n_k(1+n_k)\,\frac{1}{2}
I_{ij}(k) H_{ij} \label{def2} \\
&& f^{(1)} := - n_k(1+n_k) \phi_k\,;~~ \phi_k =\beta
\frac{1}{2}B_{ij}(\underline{k})H_{ij}  \nonumber \\
&& \underline{k}_\mu \partial^\mu f^{(1)} = \beta^2\,
n_k(1+n_k)N_k\,\frac{1}{2} I_{ij}(k) H_{ij}\,\frac{1}{2} B_{lm}(\underline{k})
H_{lm} \nonumber \\
&&f^{(2)} := n_k(1+n_k)N_k \theta_k\,;~~\theta_k:= \beta^2 \frac{1}{4}
C_{ijlm}(\underline{k}) H_{ij}H_{lm} \nonumber
\eea
where we define
\bea
\label{def3}
&& \hat I_{ij}(k) =
(\hat k_i \hat k_j-\frac{1}{3}\delta_{ij})\,;~~I_{ij}(k) = k^2\hat I_{ij}(k)
\eea
and write,
\bea
B(\underline{k})_{ij} = \hat I_{lm}(k)
B(\underline{k})\,;~~~~C_{ijlm}(\underline{k}) =  \hat I_{ij}(k) \hat I_{lm}(k)
C(\underline{k}). \label{def4}
\eea

The viscous shear stress part of the energy momentum tensor is given by
\bea
\langle  \pi_{ij}\rangle  = \int \frac{d^3 k}{(2\pi)^3 2\omega_k}  f\, (k_i
k_j-\frac{1}{3}\delta_{ij} k^2)\,.
\eea
Using the expansion (\ref{expf}) we get,
\bea \langle  \pi_{ij}\rangle  = \int \frac{d^3 k}{(2\pi)^3 2\omega_k} [f^{(0)}
+ f^{(1)} + f^{(2)}]\,(k_i k_j-\frac{1}{3}\delta_{ij}k^2)\,.
\eea
The lowest order term is identically zero.  We obtain the linear and quadratic
contributions by substituting in   (\ref{def2}) and (\ref{def4}). We use
$H_{ii}:=0$
and the following results which are obtained from rotational invariance:
\bea
&& k_i k_j B(\underline{k}) ~~ \rightarrow ~~ \frac{1}{3} \delta_{ij} k^2
B(\underline{k}) \nonumber \\
&& k_i k_j \hat k_l \hat k_m B(\underline{k}) ~~\rightarrow~~
\frac{1}{15}(\delta_{ij}\delta_{lm}+\delta_{il}\delta_{jm}+
\delta_{im}\delta_{jl}) k^2 B(\underline{k}) \nonumber \\
&& k_i k_j \hat k_l \hat k_m  \hat k_a \hat k_b C(\underline{k})
{}~~\rightarrow~~ \frac{1}{105}\left[\right.
\delta_{ab}(\delta_{lm}\delta_{ij} + \delta_{lj}\delta_{mi} +
\delta_{li}\delta_{mj})
+ \delta_{al}(\delta_{bm}\delta_{ij} + \delta_{bi}\delta_{mj} +
\delta_{bj}\delta_{mi}) \nonumber \\
&&~~~~~~~~~~~~~~~~~~~~~~~~~~~~~~~~~~~~+ \delta_{am}(\delta_{bl}\delta_{ij} +
\delta_{bj}\delta_{li} + \delta_{bi}\delta_{lj})
+ \delta_{ai}(\delta_{bl}\delta_{mj} + \delta_{bm}\delta_{lj} +
\delta_{bj}\delta_{ml}) \nonumber \\
&&~~~~~~~~~~~~~~~~~~~~~~~~~~~~~~~~~~~~+ \delta_{aj}(\delta_{bl}\delta_{mi} +
\delta_{bm}\delta_{li} + \delta_{bi}\delta_{ml})
\left.\right]k^2 C(\underline{k})
\eea
We obtain,
\bea
&& \delta\langle \pi_{ij}\rangle =  -\frac{\beta}{15} \int \frac{d^3
k}{(2\pi)^3 2\omega_k} n_k(1+n_k) k^2B(\underline{k})\,H_{ij} +
\frac{2\beta^2}{105}  \int \frac{d^3 k}{(2\pi)^3 2\omega_k} [n_k(1+n_k)N_k] k^2
 C(\underline{k}) H^{T2}_{ij}  \nonumber
\eea
Comparing with (\ref{DEFB}) we have,
\bea
&&\eta^{(1)} = \frac{\beta}{15} \int \frac{d^3 k}{(2\pi)^3 2\omega_k}
n_k(1+n_k) k^2B(\underline{k}) \label{ttf11} \\
&& \eta^{(2)} = \frac{2\beta^2}{105}  \int \frac{d^3 k}{(2\pi)^3 2\omega_k}
[n_k(1+n_k)N_k] k^2  C(\underline{k})\label{ttf}
\eea
Thus we have shown that the shear viscosity and the quadratic shear viscous
response can be obtained from the functions  $B(\underline{k})$ and
$C(\underline{k})$ respectively.  These two functions are the coefficients of
the linear and quadratic terms in the gradient expansion of the distribution
function.  In the next section we will show that these  functions can be
obtained from the first two equations in the hierarchy of equations obtained
from the gradient expansion of the Boltzmann equation.

\subsection{Expansion of the Boltzmann Equation}

The Boltzmann equation describes the evolution of the distribution function
$f(x,\underline k)$ and can be used to obtain integral equations for the
functions $B(\underline k)$ and $C(\underline k)$ defined in (\ref{def2}) and
(\ref{def4}). The
 Boltzmann equation has the form:
\bea
\underline{k}_\mu \partial^\mu f(x,\underline{k}) = {\cal C}[f] \label{BT}
\eea
where ${\cal C}[f]$ is the collision term:
\bea
{\cal C}[f] = \frac{1}{2} \int_{123} d \,\Gamma_{12\leftrightarrow 3k}[
f_1 f_2 (1+f_3)(1+f_k) - (1+f_1)(1+f_2) f_3 f_k ] \label{BTZE}
\eea
with $f_i:= f(x,\underline{p}_i)$, $f_k:=f(x,\underline{k})$.  The symbol 
$d \, \Gamma_{12\leftrightarrow 3k}$ 
represents the differential transition rate for
particles of
momentum $P_1$ and $P_2$ to scatter into momenta $P_3$ and $K$ and is given by
\bea
d\,\Gamma_{12\leftrightarrow 3k} := \frac{1}{2\omega_k} |{\cal
T}(\underline{k},\underline{p}_3,\underline{p}_2,\underline{p}_1|^2
\Pi^3_{i=1} \frac{d^3  p_i}{(2\pi)^32\omega_{p_i}} (2\pi)^4
\delta(\underline{p}_1 + \underline{p}_2 - \underline{p}_3 - \underline{k})
\eea
where ${\cal T}$ is the multiparticle scattering amplitude.
Using the expansion (\ref{expf}) and (\ref{def2}) produces a hierarchy of
equations.

\subsubsection{First Order Boltzmann Equation}

The first order equation is:
\be
\underline{k}^\mu\partial_\mu f^0 (x,\underline{k})={\cal C}[f^{(0)};f^{(1)}]
\label{foe}
\ee
where we keep terms linear in $f^{(1)}$ on the right hand side.
The left hand side gives,
\bea
\underline{k}^\mu\partial_\mu f^0 (x,\underline{k})
= - n_k(1+n_k)\beta\,\frac{1}{2} I_{ij}(k) H_{ij}
\eea
The right hand side is
\be
{\cal C}[f^{(1)}]= \frac{1}{2} \int_{123} d \,\Gamma_{12\leftrightarrow 3k}
(1+n_1)(1+n_2)n_3 n_k(\phi_k+\phi_3-\phi_1-\phi_2)
\ee
Using the definition of $\phi$ given in (\ref{def2}) and comparing the
coefficients of $H_{ij}$ on both sides of (\ref{foe}) we obtain,
\bea
I_{ij}(k)=\frac{1}{2} \int_{123} d \,\Gamma_{12\leftrightarrow 3k}
\frac{(1+n_1)(1+n_2)n_3}{1+ n_k}[B_{ij}(\underline{p}_1) +
B_{ij}(\underline{p}_2) - B_{ij}(\underline{k}) - B_{ij}(\underline{p}_3)\,]
\label{Bint}
\eea
This result is an inhomogeneous linear integral equation which can be solved
self-consistently to obtain the function $B_{ij}(\underline{k})$.

\subsubsection{Second Order Boltzmann Equation}
The second order contribution to (\ref{BT}) is,

\be
\underline{k}^\mu\partial_\mu f^{(1)} (x,\underline{k})={\cal
C}[f^{(0)};f^{(1)}; f^{(2)}]
\label{b2}
\ee
where we keep terms linear in $f^{(2)}$ and quadratic in $f^{(1)}$ on the right
hand side.
Using (\ref{def2}) the left hand side becomes,
\bea
\underline{k}^\mu \partial_\mu f^{(1)}=&&n_k(1+n_k)N_k \beta^2
\frac{1}{4}I_{ij}(k)B_{lm}(\underline{k}) H_{ij}H_{lm}\nonumber
\eea
 and the right hand side gives,
\bea
C^{(2)}[f^{(0)};f^{(1)}; f^{(2)}]=&& \frac{1}{2} \int_{123} d
\,\Gamma_{12\leftrightarrow 3k}
(1+n_1)(1+n_2)n_3n_k\nonumber
\\
&&[(N_1\theta_1+N_2\theta_2-N_3\theta_3-N_k\theta_k)\nonumber
\\
&&+\frac{1}{2}(
(N_1+N_2)\phi_1\phi_2-(N_k+N_3)\phi_3\phi_k+(N_3-N_1)\phi_1\phi_3\nonumber
\\
&&+(N_k-N_1)\phi_1\phi_k+(N_3-N_2)\phi_3\phi_2+(N_k-N_2)\phi_k\phi_2)]
\eea
Using the definitions of $\phi$ and $\theta$ given in (\ref{def2}) and
comparing the
coefficients of $H_{ij}H_{lm}$ on both sides  we obtain,
\bea
n_k && (1+n_k) N_kI_{ij}(k) B_{lm}(\underline{k}) = \frac{1}{2} \int_{123} d
\,\Gamma_{12\leftrightarrow 3k}
(1+n_1)(1+n_2)n_3n_k\nonumber
\\
&&\{ [ N_1 C_{ijlm}(\underline{p}_1) +
N_2 C_{ijlm}(\underline{p}_2) - N_k C_{ijlm} (\underline{k}) - N_3
C_{ijlm}(\underline{p}_3) ]\label{Cint}
\\
&&+\frac{1}{2}[( (N_1+N_2)B_{ij}(\underline{p}_1)B_{lm}(\underline{p}_2)
-(N_k+N_3)B_{ij}(\underline{p}_3)B_{lm}(\underline{k})
+(N_3-N_1)B_{ij}(\underline{p}_1)B_{lm}(\underline{p}_3) \nonumber
\\
&&+(N_k-N_1)B_{ij}(\underline{p}_1)B_{lm}(\underline{k})+
(N_3-N_2)B_{ij}(\underline{p}_3) B_{lm}(\underline{p}_2)
+(N_k-N_2)B_{ij}(\underline{k})B_{lm}(\underline{p}_2)    ] \}\nonumber
\eea
This integral equation can be solved self consistently for the quantity
$C_{ijlm}(\underline{k})$ using the result for $B_{ij}(\underline{k})$ from
(\ref{Bint}).

\section{Viscosity from Field Theory}

The Kubo formulae allow us to use quantum field theory to calculate
nonequilibrium transport coefficients.  The results should be the same as those
obtained in the previous section using transport theory.  To do calculations in
a system that is out of equilibrium, the Hamiltonian is separated into an
equilibrium piece $H_0$ and a nonequilibrium piece $H_{ext}$ which depends on
the gradients of the thermodynamic parameters: the four-velocity field and the
inverse temperature field.  For systems close to equilibrium the nonequilibrium
piece of the Hamiltonian can be treated as a perturbation and the deviations of
physical quantities from their equilibrium values can be calculated
perturbatively.  The linear response calculation includes only the first order
contribution to this expansion and gives transport coefficients that can be
expressed as integrals of retarded two-point Green functions over space and
time.
One of the results we obtain in this paper is that the quadratic shear viscous
response coefficient
can be written in terms of a retarded three-point function.  Throughout this
section we use capital letters to denote four-vectors and small letters for
three-vectors. We also define $\int d^4p/(2\pi)^4:= \int dP $.

To calculate the expectation value of an operator we take the trace over the
density matrix.  We follow the presentation of \cite{Hosoya}. We work with the
density matrix in the Heisenberg representation which satisfies,

\be
\frac{\partial \rho}{\partial t}=0
\ee
and can be written as,

\be
\rho=\frac{e^{-A+B}}{{\rm Tr} e^{-A+B}     }
\ee
where
\bea
A &=&\int d^3 x F^\nu T_{0\nu},\nonumber \\
B&=&\int d^3 x \int _{-\infty}^t dt' e^{\epsilon (t'-t)}T_{\mu\nu}(x,t')
\partial^\mu F^\nu(x,t')
\label{AB}
\eea
with $F^\mu=\beta u^\mu$ and $\epsilon$ to be taken to zero at the end.  In
this expression $A$ is the equilibrium part of the Hamiltonian and $B$ is a
perturbative contribution that is linear in the gradient of the four-velocity
field. Note that in the local rest frame  $u^\mu = (1,0,0,0)$ and 
$[A]^{LRF} = \int d^3x\,\beta T_{00}$.
Using the identity for the exponential function of two operators
\be
e^{\beta(\hat a+\hat b)}=e^{\beta  \hat a}[1+\int_0^\beta d\lambda
e^{\lambda\hat a} \hat b e^{-\lambda(\hat a+\hat b)}]
\ee
we can expand
\bea
e^{-A+B}=&&e^{-A}[1+\int_0^1 d\lambda e^{\lambda A} B e^{-\lambda A}\nonumber
\\
&&
+\int_0^1 d\lambda \int_0^\lambda d\tau e^{\lambda A} B e^{-\lambda A}
e^{\tau A} B e^{-\tau A}+ {\cal O}(B^3)]
\eea
 Thus the density matrix can be written as
\bea
\rho=&&\rho_0 \left[ \right. 1+\int_0^1d\lambda( B(\lambda) -\langle
B(\lambda)\rangle ) +\int_0^1 d\lambda \int_0^\lambda d\tau (B(\lambda)
B(\tau)-\langle B(\lambda) B(\tau)\rangle  )
\nonumber
\\
&&-\int_0^1 d\lambda \int_0^1 d\lambda'(\langle  B(\lambda)\rangle
B(\lambda')-\langle  B(\lambda)\rangle \langle  B(\lambda')\rangle
\left.\right] + O(B^3)
\label{exrho}
\eea
where
\be
\rho_0=\frac{e^{-A}}{{\rm Tr} e^{-A}}
\ee
is the local equilibrium density matrix.

\subsection{Viscosity as an Expansion in Green Functions of Composite
Operators}

\subsubsection{Linear Response}

Using the first three terms of (\ref{exrho}) produces the linear response
approximation for the deviation of the expectation value of the viscous shear
stress part of the energy momentum tensor from the equilibrium value:
\bea
\delta \langle  \pi_{\mu\nu}(x,t)\rangle^l  &&= \langle
\pi_{\mu\nu}(x,t)\rangle^l  - \langle  \pi_{\mu\nu}(x,t)\rangle _0 \nonumber \\
&& = \int d^3 x' \int_{-\infty}^{t} dt'e^{\epsilon (t'-t)} \bigl(
\pi_{\alpha\beta}(x',t'), \pi_{\mu\nu}(x,t) \bigr)\partial^\alpha
F^\beta(x',t'')
\eea
where the correlation function
$\bigl( \pi_{\alpha\beta}(x',t'), \pi_{\mu\nu}(x,t)\bigr) $ is defined as,

\be
\bigl( \pi_{\mu\nu}(x,t)
, \pi_{\alpha\beta}(x',t')\bigr)=\frac{1}{\beta}
\int_0^\beta d\tau
\la(\pi_{\alpha\beta}(x',t'+i\tau) \pi_{\mu\nu}(x,t)-\la
\pi_{\alpha\beta}(x',t'+i\tau)
\rangle_0 \pi_{\mu\nu}(x,t)\rangle_0 \ee
and $\langle  \cdots\rangle _0 := {\rm Tr}(\rho_0 \cdots )$. 
From now on we drop the
subscript $0$ on the
correlation functions.  We
take the initial time $t_0$ to
minus infinity and we assume
the system is in equilibrium
at $t=t_0$ and that the
external forces are switched
on adiabatically. We
assume that changes in the
thermodynamic forces are
 small enough over the
correlation length of the
correlation functions that
factors $F_\mu = \beta u_\mu$
can be taken out of the
integral.   Using (\ref{def1})
and rotational invariance we
have,
\bea
\pi_{\mu\nu}\pi_{\alpha\beta}
=\frac{1}{10}(\Delta_{\mu\alpha}\Delta_{\nu\beta}
+\Delta_{\mu\beta}\Delta_{\nu\alpha}-\frac{2}{3}\Delta_{\mu\nu}
\Delta_{\alpha\beta})
\pi_{\sigma\tau}\,
\pi^{\sigma\tau}  \eea Using
this result we obtain, \bea
\delta \langle \pi_{\mu\nu}
\rangle^l&&=\frac{H_{\mu\nu}}{
10}\int d^3 x'
\int_{-\infty}^{t}
dt'e^{\epsilon (t'-t)}
\nonumber \\ &&~~\int_0^\beta(
\la(\pi(x',t+i\tau)
\pi(x,t)\rangle-\la\la
\pi(x',t+i\tau)\rangle\pi (x,t)\rangle)\eea
where we have written $\pi_{\sigma\tau} \pi^{\sigma\tau} := \pi \pi$. If we
assume that correlations vanish as $t'\to -\infty$ this expression 
can be rewritten as,
\bea \delta
\langle \pi_{\mu\nu} \rangle^l
&&= \frac{H_{\mu\nu}}{10}\int
d^3 x' \int_{-\infty}^{t}
dt'e^{\epsilon
(t'-t)}\nonumber \\ &&
{}~~\int_0^\beta
\int_{-\infty}^{t'}d t''
\frac{d}{d t''} (
\la(\pi(x',t''+i\tau)
\pi(x,t)\rangle-\la\la
\pi(x',t''+i\tau)\rangle\pi
(x,t)\rangle) \eea   
Using
$\frac{\partial}{dt''}f(t''+i\tau) =-i\frac{\partial}{\partial \tau} 
f(t''+i\tau)$ we can
perform the integration over $\tau$.  We obtain,
\bea
&& \delta \langle\pi_{\mu\nu}\rangle^l  \\
&&~~= -\frac{i}{10} H_{\mu\nu} \int d^3 x' \int_{-\infty}^{t}
dt'e^{\epsilon (t'-t)} \int_{-\infty}^{t'}d t'' ( \langle \pi(x',t''+i\beta) \pi(x,t)\rangle
-\langle  \pi(x',t'')\pi(x,t)\rangle ) \nonumber \,.
\eea
Using the KMS condition
\bea
\langle \pi(x',t''+i\beta) \pi(x,t)\rangle &&= {\rm Tr} [e^{-\beta H}
\pi(x',t''+i\beta)\pi (x,t)] \nonumber \\
&&=
{\rm Tr} [e^{-\beta H} \pi(x,t)\pi (x',t'')] = \la \pi(x,t)\pi (x',t'') \rangle
\eea
we obtain,

\be
\delta \langle  \pi_{\mu\nu}\rangle^l  = \frac{H_{\mu\nu}}{10}\int d^3 x'
\int_{-\infty}^{t} dt'e^{\epsilon (t'-t)} \int_{-\infty}^{t'}d
t''D_R(x,t;x',t'')
\ee
where
\be
D_R(x,t;x',t'')=-i\theta(t-t'')[\pi(x,t),\pi(x',t'')]
\ee
We extract the shear viscosity using the definition (\ref{DEF}).
We obtain,
\bea
\eta^{(1)} = \frac{1}{10} \int d^3 x' \int_{-\infty}^{0} dt'e^{\epsilon (t'-t)}
\int_{-\infty}^{t'}d t''D_R(0;x',t'')
\eea
We can rewrite this result in a more useful form.
We rewrite the integrals inserting theta functions:
$\int^0_{-\infty} dt' = \theta(-t') \int^\infty _{-\infty}dt'$ and use the
integral representation for the theta function:
\bea
\theta(x) = \int \frac{d\omega}{2\pi i} \frac{e^{i\omega x}}{\omega
-i\epsilon}. \label{theta}
\eea
Inserting $D_R(0;x',t'') = \int dQ \,e^{iQX'} D_R(Q)$ with $X'=(x',t'')$  we
obtain,
\bea
\eta^{(1)} = -\frac{i}{10}\frac{d}{d q_0}[\lim_{\vec{q} \to 0}D_R(Q)]|_{q_0=0}
\eea
Since it is the imaginary part of the two-point function that is odd in $q_0$
we have,
\bea
\eta^{(1)} = \frac{1}{10}\frac{d}{d q_0}{\rm Im} [ \lim_{\vec{q} \to
0}D_R(Q)]|_{q_0=0} \label{pl}
\eea
This is the well known Kubo formula which expresses the shear viscosity in
terms of a retarded two-point green function that can be calculated using
equilibrium quantum field theory.

\subsubsection{Quadratic Response}

Now we consider corrections to the linear response approximation.  We calculate
the quadratic shear viscous response from the terms in (\ref{exrho}) that are
quadratic in the interaction. We show that the result can be written as a
retarded three-point correlator. We obtain,
\bea
\delta \langle  \pi_{\mu\nu}(x,t)\rangle ^{q}
=&&\int d^3 x' \int d^3 x'' \int_{-\infty}^{t} dt'e^{\epsilon (t'-t)}
\int_{-\infty}^{t} dt''e^{\epsilon (t''-t)}\nonumber
\\
&&
\bigl( \pi_{\mu\nu}(x,t)
, \pi_{\alpha\beta}(x',t'),\pi_{\rho\sigma}(x'',t'')\bigr)\partial^\alpha
F^\beta(x',t')
\partial^\rho F^\sigma(x'',t'')\,.
\eea
We use (\ref{def1}) and rotational invariance to write,
\bea
\pi_{\mu\nu}(x,t)
 \pi_{\alpha\beta}(x',t')\pi_{\rho\sigma}(x'',t'') \partial^\alpha u^\beta
\partial^\lambda u^\tau && = \frac{3}{35} H^{T2}_{\mu\nu} (
\pi_{\alpha\beta}(x,t)\pi^{\beta\lambda}(x',t')\pi^{~\alpha}_\lambda
(x'',t'') )\nonumber \\
&&:= \frac{3}{35} H^{T2}_{\mu\nu} ( \pi(x,t)\pi(x',t')\pi(x'',t'') )
\eea
Writing out the correlation function we obtain,
\bea
\delta \langle  \pi_{\mu\nu}(x,t)\rangle ^{q}
=&& \frac{3}{35} H_{\mu\nu}^{T2}
\int d^3 x' \int d^3 x'' \int_{-\infty}^{t} dt'e^{\epsilon (t'-t)}
\int_{-\infty}^{t} dt''e^{\epsilon (t''-t)}\nonumber\\
&&
\left[\right.\int_0^\beta d\tau\int_0^\tau d\lambda \langle  \pi(x',t'+i\tau)
\pi(x'',t''+i\lambda) \pi(x,t)\rangle \nonumber
\\
&&-
\int_0^\beta d\tau\int_0^\beta d\tau' \langle \langle  \pi(x',t'+i\tau)\rangle
\pi(x'',t''+i\tau') \pi(x,t)\rangle \nonumber
\\
&&-
\int_0^\beta d\tau\int_0^\tau d\lambda \langle \langle  \pi(x',t'+i\tau)
\pi(x'',t''+i\lambda)\rangle  \pi(x,t)\rangle \nonumber
\\
&&+\int_0^\beta d\tau\int_0^\beta d\tau' \langle \langle
\pi(x',t'+i\tau)\rangle
\langle \pi(x'',t''+i\tau')\rangle  \pi(x,t)\rangle \left.\right] \label{int1}
\eea
Assuming that correlations vanish when the time approaches minus infinity we
can rewrite
(\ref{int1}) as
\bea
\delta \langle  \pi_{\mu\nu}(x,t)\rangle  ^{q}
=&&\frac{3}{35} H_{\mu\nu}^{T2} \int d^3 x' \int d^3 x'' \int_{-\infty}^{t} dt'
e^{\epsilon (t'-t)}
\int_{-\infty}^{t} dt''e^{\epsilon (t''-t)}\int_{-\infty}^{t'}
ds'\int_{-\infty}^{t''} ds''
\nonumber\\
&&
\frac{d}{ds'}\frac{d}{ds''} \left[ \right.\int_0^\beta d\tau\int_0^\tau
d\lambda  \langle  \pi(x',s'+i\tau)
\pi(x'',s''+i\lambda) \pi(x,t)\rangle \nonumber
\\
&&-
\int_0^\beta d\tau\int_0^\beta d\tau' \langle \langle  \pi(x',s'+i\tau)\rangle
\pi(x'',s''+i\tau') \pi(x,t)\rangle \nonumber
\\
&&-
\int_0^\beta d\tau\int_0^\tau d\lambda \langle \langle  \pi(x',s'+i\tau)
\pi(x'',s''+i\lambda)\rangle  \pi(x,t)\rangle \nonumber
\\
&&+\int_0^\beta d\tau\int_0^\beta d\tau' \langle \langle
\pi(x',s'+i\tau)\rangle
\langle \pi(x'',s''+i\tau')\rangle  \pi(x,t)\rangle \left.\right]
\eea
 Carrying out the integration over $\lambda$, $\tau$ and $\tau'$ and using the
fact that we have symmetry under interchange of $(x',t')$ and $(x'',t'')$ we
obtain,
\bea
\delta \langle  \pi_{\mu\nu}(x,t)\rangle ^{q}
=&&\frac{3}{70} H_{\mu\nu}^{T2} \int d^3 x' \int d^3 x'' \int_{-\infty}^{t}
dt'e^{\epsilon (t'-t)}
\int_{-\infty}^{t} dt''e^{\epsilon (t''-t)}\int_{-\infty}^{t'}
ds'\int_{-\infty}^{t''} ds''\nonumber
\\
&&
\frac{1}{2}([[\pi(x,t),\pi(x',s')],\pi(x'',s'')]+
[[\pi(x,t),\pi(x'',s'')],\pi(x',s')])
\eea
We define
\bea
G(x,t;x',s';x'',s'') =
\frac{1}{2}([[\pi(x,t),\pi(x',s')],\pi(x'',s'')]+
[[\pi(x,t),\pi(x'',s'')],\pi(x',s')])
\eea
and write
\bea
&&\delta \langle \pi_{\mu\nu}(x,t) \rangle ^{q} \nonumber \\
&&= \frac{3}{70} H_{\mu\nu}^{T2} \int d^3 x' \int d^3 x'' \int_{-\infty}^{t}
dt'e^{\epsilon (t'-t)}
\int_{-\infty}^{t} dt''e^{\epsilon (t''-t)}\int_{-\infty}^{t'}
ds'\int_{-\infty}^{t''} ds''G(x,t;x',s';x'',s'') \nonumber
\eea
Using (\ref{DEF}) we extract,
\bea
\eta^{(2)} = \frac{3}{70} \int d^3 x' \int_{-\infty}^{0} dt'e^{\epsilon t'}
\int_{-\infty}^{0}d t''e^{\epsilon t''} \int_{-\infty}^{t'} ds'
\int_{-\infty}^{t''} ds'' G(0;x',s';x'',s'')
\eea
As we did previously, we can rewrite this result in a neater form.
We rewrite the integrals inserting theta functions:
$\int^0_{-\infty} dt' = \theta(-t') \int^\infty _{-\infty}dt'$ and use the
integral representation for the theta function (\ref{theta}).
We use $X'=(x',s')$ and $X'' = (x'',s'')$ and rewrite,
\bea
G(0;x',s';x'',s'') = \int dQ e^{iX'Q}\int dQ' e^{i X''Q'} G(-Q-Q',Q,Q')\,.
\eea
We obtain,
\bea
\eta^{(2)} = \frac{3}{70}\frac{d}{d q_0} \frac{d}{ dq_0'} [ \lim_{\vec{q} \to
0}G(-Q-Q',Q,Q')]|_{q_0=q'_0=0}
\eea
Using the techniques of \cite{mu} to write the three-point function in the
spectral representation, it is tedious but straightforward to show that the
only contribution to the result above comes from the real part of the
three-point function which is retarded with respect to the first leg.  In
coordinate space this retarded three-point function is written: $G_{R1}(x,y,z)
= \theta(t_x-t_y)\theta(t_y-t_z)[[\pi(x),\pi(y)],\pi(z)] +
\theta(t_x-t_z)\theta(t_z-t_y)[[\pi(x),\pi(z)],\pi(y)]$. We obtain,
\bea
\eta^{(2)} = \frac{3}{70}\frac{d}{d q_0} \frac{d}{ dq_0'}{\rm Re}\,[
\lim_{\vec{q} \to 0}G_{R1}(-Q-Q',Q,Q')]|_{q_0=q'_0=0} \label{pq}\,.
\eea
This is an interesting new result.  We have obtained a type of nonlinear Kubo
formula that allows us to obtain the quadratic shear viscous response from a
retarded three-point function using equilibrium quantum field theory.

\subsection{Diagrammatic Expansion}

We obtain a diagrammatic expansion for the viscosity coefficients given in
(\ref{pl}) and (\ref{pq}).    We use the closed time path formulation of finite
temperature field theory, and work in the Keldysh representation.  Several
reviews of this technique are available in the literature
\cite{keld,schw,Chou,rep-145,PeterH}.  The closed time path integration contour
involves two branches, one running from minus infinity to
positive infinity just above the real axis, and one running back from positive
infinity to
negative infinity just below the real axis.  All fields can take values on
either branch of
the contour and thus there is a doubling of the number of degrees of freedom.
It is straightforward to show that this doubling
of degrees of freedom is necessary to obtain finite green functions. We discuss
below the structure of correlation functions of field operators.

The two-point function or the propagator can be written as a $2 \times 2 $
matrix of the form
 \begin{equation}
 \label{2}
   D = \left(  \matrix {D_{11} & D_{12} \cr
                        D_{21} & D_{22} \cr} \right) \, ,
 \end{equation}
where $D_{11}$ is the propagator for fields moving along $C_1$,
$D_{12}$ is the propagator for fields moving from $C_1$ to $C_2$, etc.
The four components are given by
 \begin{eqnarray}
 \label{eq: compD}
   D_{11}(X-Y) &=& -i\langle T(\phi(X) \phi(Y))\rangle  \nonumber\\
   D_{12}(X-Y) &=& -i\langle \phi(Y) \phi(X) \rangle  \nonumber\\
   D_{21}(X-Y) &=& -i\langle \phi(X) \phi(Y)\rangle  \nonumber\\
   D_{22}(X-Y) &=& -i\langle\tilde{T}(\phi(X)\phi(Y))\rangle
 \end{eqnarray}
where $T$ is the usual time ordering operator and $\tilde{T}$ is the
anti-chronological time ordering operator. Physical functions are obtained by
taking appropriate combinations of the components of the propagator matrix.  It
is straightforward to show that the usual retarded and advanced propagators:
$D_R = -i\theta(x_0-y_0)[\phi(X),\phi(Y)]$ and 
$D_A =-i\theta(y_0-x_0)[\phi(X),\phi(Y)]$ are given by the combinations,
\bea \label{3a}
   D_R &=& D_{11} - D_{12}  \nonumber\\
   D_A &=& D_{11} - D_{21} \, .
\eea
The 1PI part of the two-point function, or the polarization insertion, is
obtained by truncating legs.  The retarded and advanced parts are given by,
\bea
 \Pi_R &=& \Pi_{11} + \Pi_{12}  \nonumber\\
 \Pi_A &=& \Pi_{11} + \Pi_{21} \, .
\eea

The situation is similar for higher $n$-point functions.  For example, the
three-point function which is retarded with respect to the first leg is given
by
\bea
\Gamma_{R1} = \Gamma_{111} +\Gamma_{112} +\Gamma_{121} +\Gamma_{122}\,.
\eea
The other three-point vertices that we will  need are:
\bea
&& \Gamma_{R2} = \Gamma_{111} +\Gamma_{112} +\Gamma_{211} +\Gamma_{212}
\nonumber \\
&& \Gamma_{R3} = \Gamma_{111} +\Gamma_{121} +\Gamma_{211} +\Gamma_{221}
\nonumber \\
&& \Gamma_{F} = \Gamma_{111} +\Gamma_{121} +\Gamma_{212} +\Gamma_{222}
\nonumber 
\eea
The four-point function which is retarded with respect to the first leg is
given by,
\bea
M_{R1} = M_{1111} +M_{1112} +M_{1121} +M_{1211} + M_{1122} +M_{1212} +M_{1221}
+M_{1222} \,.
\eea
The other four-point vertices that we will need are:
\bea
&& M_{R4} = M_{1111} +M_{1121} +M_{1211} +M_{2111} + M_{1221} +M_{2121}
+M_{2211} +M_{2221}\nonumber\\
&& M_F = M_{1111} +M_{1121} +M_{1211} +M_{1221}  +M_{2112} + M_{2212} +M_{2122}
+M_{2222} \,.
\eea
There are two relations that we can use to simplify expressions involving these
vertices.  The first is a consequence of the fact that real time green
functions are related to each other through the KMS conditions.  These
identities are a consequence of symmetries generated through cyclic
permutations of field operators.  The identity that we will need is \cite{mu},
\bea
\Gamma_F + N_1 \Gamma_{R3} + N_3 \Gamma_{R1} = (N_1 + N_3) \Gamma^*_{R2}
\label{kms}
\eea
where each vertex carries arguments $\Gamma (P_1,P_2,P_3)$ and we write
$N(P_1):=N_1$ etc.
The second simplification occurs because of the fact that the equations we
obtain contain a specific combination of four-point vertices which we define
as,
\bea
\bar M_F = M_F + N_1 M_{R4} + N_4 M_{R1} \label{mfbar}
\eea
where each vertex carries arguments $M (P_1,P_2,P_3,P_4)$.
It has been shown that this particular combination of four-point vertices has
special properties in a number of other contexts \cite{johnS}.

In addition, bare vertices that carry a Keldysh index `2' have an extra factor
of minus one.  This factor is accounted for by inserting a factor of $\tau$ for
each vertex where $\tau$ is the two component vector given by,
\bea
\tau = {1 \choose -1} \label{tau} \,.\nonumber
\eea

We want to obtain a perturbative expansion for the correlation functions of
composite operators
 $D_R(X,Y)$ and $G_{R1}(X,Y,Z)$ which appear in (\ref{pl}) and (\ref{pq}).  We
obtain expressions for these correlation functions in terms of the vertices
$\Gamma_{ij}$ and $M_{ijlm}$ which are defined as the vertices obtained by
truncating external legs from the following connected vertices:
\bea
&& \Gamma^C_{ij} = \langle T_c \pi_{ij}(X) \phi(Y) \phi(Z)\rangle  \nonumber \\
&& M^C_{ijlm} = \langle T_c \pi_{ij}(X) \pi_{lm}(Y)\phi(Z) \phi(W)\rangle
\eea
where
\bea
\pi_{ij}(X) = \partial_i\phi(X) \partial_j\phi(X) - \frac{1}{3} \delta_{ij}
(\partial_m \phi(X))(\partial_m \phi(X))
\eea
and $T_c$ is the operator that time orders along the closed time path contour.
These definitions allow us to write the two- and three-point correlation
functions as integrals of the form,

%%%%%%%%%%%%%%%%%%%%%%%%%%%%% Fig. 1 %%%%%%%%%%%%%%%%%%%%%%%%%%%%%%%%%%%%%
\begin{eqnarray}
\parbox{14cm}
{{
\begin{center}
\parbox{10cm}
{
\epsfxsize=8cm
\epsfysize=8cm
\epsfbox{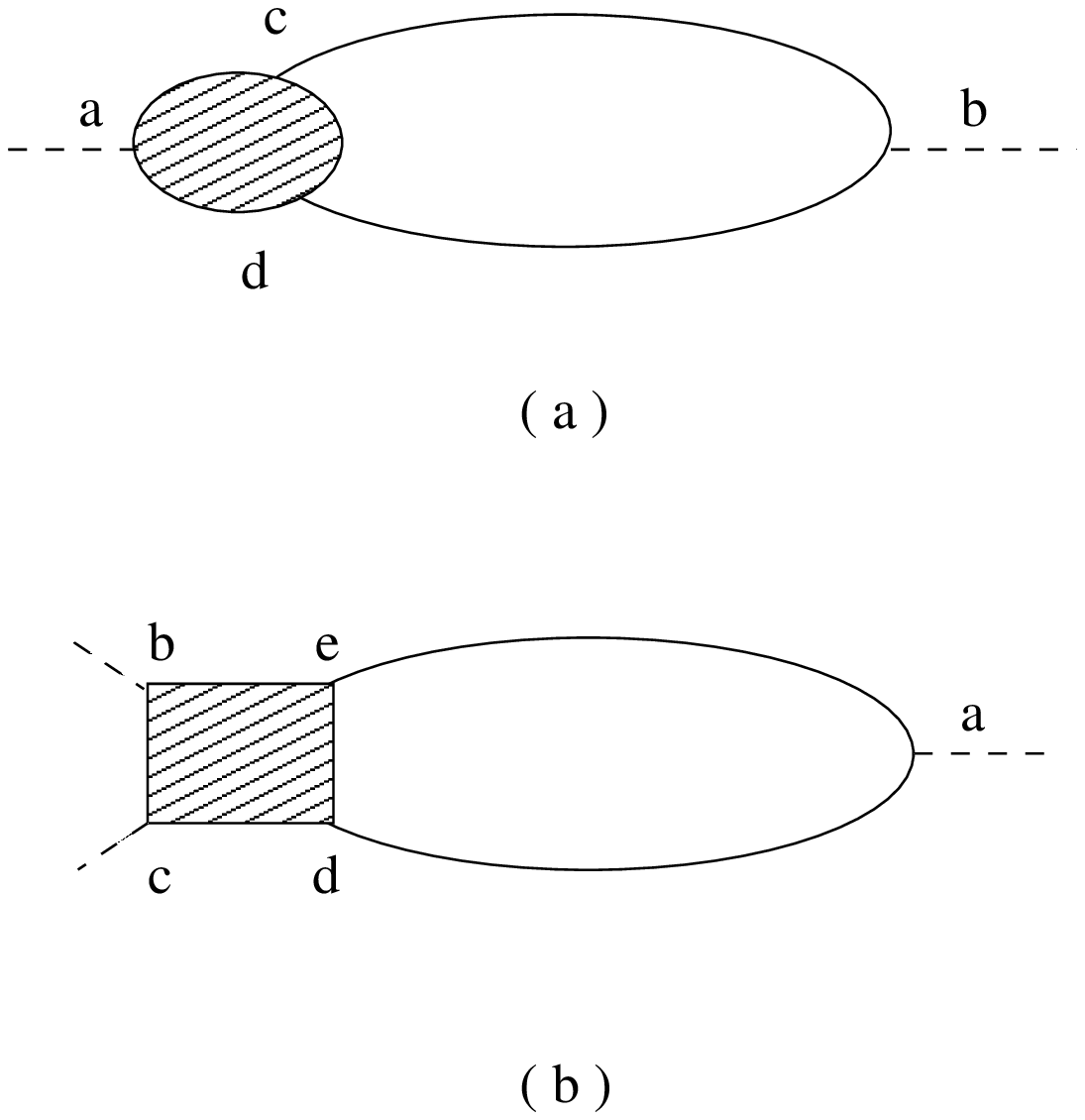}}\\
\parbox{14cm}{\small \center  Fig.~[1]: (a) Two-point function
 for shear viscosity from linear response; (b) Three-point function for
quadratic shear viscous response. The dashed external line represents the
composite operator $\pi_{ij}$. The square box is the four-point function $M$
and the round blob is the three-point vertex $\Gamma$.  }

%%%%%%%%%%%%%%%%%%                           %%%%%%%%%%%%%%%%%%%%%%%%%%
%%%%%%%%%%%%%%%%%%%%%%%%%%%%%%%%%%%%%%%%%%%%%%%%%%%%%%%%%%%%%%%%%%%%%%%

%\label{F2}
\end{center}
}}
\nonumber
\end{eqnarray}
\bea
&& D_{ab}(Q) = 2i \int dP \, \Gamma^{ij}_{cad}(P,Q,-P-Q) iD_{bc}(P) iD_{db}(P+Q)
I_{ji}(p) \tau_b  \\
&& G_{abc}(-Q-Q',Q,Q')  =8 \int dK\, M^{ijjk}_{ebcd}\tau_a iD_{ae}(K)
iD_{da}(K+Q+Q') I_{ki}(k) \nonumber
\eea
where the indices $\{a,b,c,d,e\}$ are Keldysh indices and take values
$\{1,2\}$. These expressions are shown diagrammatically in Fig.~[1]. We perform
the sum over Keldysh indices using the Mathematica program described in
\cite{johnS}.
We obtain,
\bea
&& D_R(Q) = i\int dK (N_{k+q} - N_k) \Gamma_{R2}^{ij}(K,Q,-K-Q) D_A(K) D_R(K+Q)
I_{ji}(k) \label{kubo1} \\
&& G_{R1}(-Q-Q',Q,Q') \label{kubo2} \\
&&~~~~~~~~=  -4\int dK (\bar M_F)_{ikkj}(K,Q,Q',-K-Q-Q') D_R(K) D_A(K+Q+Q')
I_{ji}(k)  \nonumber
\eea
 To rewrite these expressions in a simpler form we use rotational invariance to
write,
\bea
M_{ijlm} := \hat I_{ij} \hat I_{lm} M\,;~~~~\Gamma_{ij} := \hat I_{ij} \Gamma
\label{def5}
\eea
Using an obvious notation we  write the pairs of propagators $D_A(P) D_R(P+Q)$
and $D_R(K) D_A(K+Q+Q')$ as $a_ p r_{p+q}$ and $r_k a_{k+q+q'}$.  When
$\{q_0\,,~q'_0 \}\rightarrow 0$ the dominant contribution to the integral from
these pairs of propagators is produced by what is called the pinch effect:  the
contour is ``pinched'' between the poles of the two propagators
which gives rise to a factor in the denominator that is proportional to the
imaginary part of the propagators.  We regulate the pinching singularity with
the imaginary part of the hard thermal loop self energy and obtain \cite{meg},
\bea
r_k a_{k+q} \rightarrow  -\frac{\rho_k}{2\rm{Im} \Sigma_k}\,;~~~~\rho_k =
i(r_k-a_k) \label{pinch}
\eea
where $\Sigma$ is the retarded part of the hard thermal loop self energy.

Now we expand in $q_0$ and $q'_0$. In (\ref{kubo1}) we keep terms proportional
to $q_0$ since these terms are the only ones that contribute to (\ref{pl}); in
(\ref{kubo2}) we keep terms proportional to $q_0 q'_0$ since these terms are
the only ones that contribute to (\ref{pq}). In each term there are products of
thermal factors of the form $N_x-N_{x+q}$ and $N_x-N_{x+q'}$ where $x$ is some
combination of $\{k,\,p,\,r\}$. The expansion of these thermal factors is
straightforward:
\bea
N_x -  N_{x+q} = 2q_0 \beta n_x(1+n_x) + \cdots \label{nexp}
\eea

Consider the behaviour of the vertices when 
$\{q_0,~q_0'\}\rightarrow 0$. Using (\ref{kubo1}) and (\ref{pinch}) 
in (\ref{pl}) it is easy to see that
only the real part of $\Gamma_{R2}$ contributes.  Similarly, using
(\ref{kubo2}) and (\ref{pinch}) in (\ref{pq}) it is clear that we need only the
real part of $\bar M_F$.  First we consider (\ref{kubo1}).  Because of the
explicit factor of $(N_{k+q} - N_k)$ we can set $Q$ to zero in the vertex.  The
expansion of (\ref{kubo2}) in $Q$ and $Q'$ is more difficult.  One can show that the
terms in the expansion that contain gradients acting on the vertices
$M_F,~M_{R1}$ and $M_{R4}$ do not contribute to the result and that we can
make the replacement:
\bea
{\rm Re} \bar M_F(K,Q,Q',-K-Q-Q') \rightarrow  (N_k - N_{k+q+q'}){\rm Re}
M_{R1}(K,0,0,-K) \label{re2}
\eea
{}From now on, to simplify the notation, we define $\Gamma(K,0,-K):=\Gamma(K)$
and $M(K,0,0,-K):=M(K)$.

Using these results to simplify (\ref{kubo1}) and (\ref{kubo2}) and
substituting into (\ref{pl}) and (\ref{pq}) we obtain,
\bea
&& \eta^{(1)} = \frac{\beta}{15} \int dK\, k^2 \,\rho_k n_k (1+n_k) \left[\frac
{\rm Re \Gamma_{R2}(K)}{\rm Im \Sigma_k}\right] \label{ftf11} \\
&& \eta^{(2)} =  -\frac{2\beta^2}{105} \int dK \,k^2\,\rho_k n_k (1+n_k) N_k
\left[ \frac{\rm Re M_{R1}(K)}{\rm Im \Sigma_k} \right]\label{ftf}
\eea
Comparing with (\ref{ttf11}) and (\ref{ttf}) we see that the results are
identical if we identify
\bea
&&B(\underline{k}) = \frac {\rm Re \Gamma_{R2}(\underline{k})}{\rm Im \Sigma_k}
\label{EQ1}\\
&& C(\underline{k}) = -  \frac{\rm Re M_{R1}(\underline{k})}{\rm Im \Sigma_k}
\label{EQ2}
\eea
%%%%%%%%%%%%%%%%%%%%%%%%%%%%%%%%%
with the momentum $K$ on the shifted mass shell: $\delta(K^2-m_{th}^2)$ where
$m_{th}^2=m^2+{\rm Re} \Sigma_K$.

 %%%%%%%%%%%%%%%%%%%%%%%%%%%%%%%%%
%%%%%%%%%%%%%%%%%%%%%%%%%%%%%%%%%%%%%%%%%%%%%%%
%%%%%%%%%%%%%%%%%%%%%%%%%%%%%%%%%%%%%%%

\subsubsection{The Ladder Resummation}

It has been known for some time that the set of diagrams which give the
dominate contributions to the vertex $\Gamma_{ij}(P,Q,-P-Q)$ are the ladder
diagrams.
These diagrams contribute to the viscosity to the same order in perturbation
theory as the
bare one loop graph and thus need to be resummed \cite{jeon1}. This
effect occurs for the following reason. It appears that the ladder graphs are
suppressed relative to the one loop graph by extra powers of the coupling
which come from the extra vertex factors that one obtains when one adds rungs
(vertical lines). However, these extra factors of the coupling are compensated
for by a kinematical factor.  This factor arises through the pinch effect which
is described above.  The addition of an additional rung in a ladder graph
always produces an extra pair of propagators of the form $a_k r_{k+q}$ or 
$r_k a_{k+q}$.
Products of this form contribute  a factor which produces an enhancement.  This
factor
occurs when the contour is ``pinched'' between the poles of the two propagators, 
which gives rise to a contribution in the denominator that is proportional to
the
imaginary part of the inverse propagators.

In order to include ladder diagrams we obtain the vertex 
$\tilde \Gamma_{ij}(P,Q,-P-Q)$ as the solution to the integral 
equation shown in Fig.~[2].

%%%%%%%%%%%%%%%%%%%%%%%%%%%%%%%%%%FIG2
\begin{eqnarray}
\parbox{14cm}
{{
\begin{center}
\parbox{10cm}
{
\epsfxsize=8cm
\epsfysize=3cm
\epsfbox{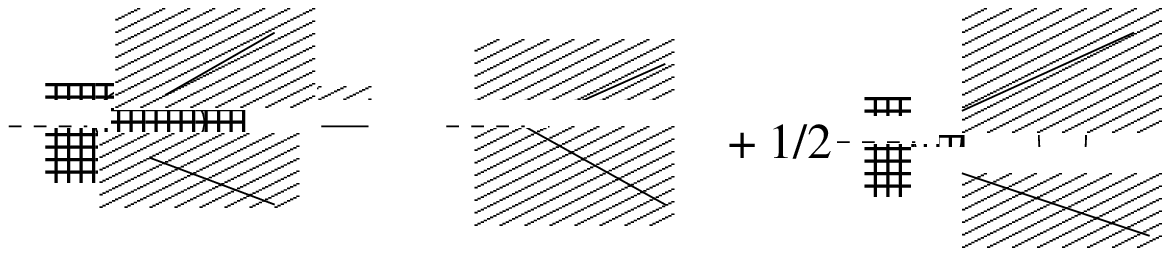}}\\
\parbox{14cm}{\small \center  Fig.~[2]: Integral equation for the ladder
resummation. The blob represents the vertex  $\tilde \Gamma$. }
\label{F02}
\end{center}
}}
\nonumber
\end{eqnarray}
\bea
&& \tilde \Gamma^{lm}_{abc}(K,Q, -K-Q) \nonumber \\
&&=  \int dP\,dR\,\tau_3^b \tau_3^a\tau_3^c I_{lm}(p) D_{ca}(R) D_{ac}(K+R-P)
D_{bc}(P+Q) D_{ab}(P) \\
&&+ \frac{1}{2} \int dP\,dR\, \tilde \Gamma^{lm}_{dbe}(P,Q,-P-Q)  D_{ad}(P)
D_{ec}(P+Q) D_{ac}(R+K-P) \tau_3^c D_{ca}(R) \tau_3^a \nonumber
\eea
We perform the sums over the Keldysh indices using the Mathematica program in
\cite{johnS} and simplify the result by taking $Q$ to zero, keeping only the
pinching terms, and using (\ref{pinch}).  The vertex function which includes the tree vertex is obtained by
shifting: 
$ \Gamma_{ij} = I_{ij} + \frac{1}{2}\tilde \Gamma_{ij}$. 
(The factor of 1/2 is a symmetry factor).  We obtain,
\bea
\label{gammaintsh1}
\Gamma_{R2}^{lm}(K) = &&k_m k_l -\frac{1}{3}
 \delta_{ml} k^2 - \frac{\lambda^2}{4}\int dP\,dR\,dP'\,\nonumber \\
&& \cdot (2\pi)^4 \delta^4(P+P'-R-K)
\rho_p \rho_r\rho_{p'} \frac{\Gamma_{R2}^{lm}(P)}{{\rm Im}\Sigma_R(P)}
(1+n_p)(1+n_{p'})n_r/(1+n_k)
\eea
Note that this integral equation is decoupled: the only three-point vertex that
appears is $\Gamma^{lm}_{R2}$. To simplify this expression further we use (\ref{def5}), (\ref{EQ1}) and the fact that
$\Gamma_{R2}(P)$ is pure real, and symmetrize the integral on the right hand
side over the integration variables $\{P,\, P',\, R\}$. We multiply and divide
the left hand side by $\rm{Im} \Sigma_k$ and replace this expression in the
numerator by the HTL result
\cite{jeon1,par},
\bea
{\rm Im}\Sigma_k =&&-\frac{\lambda^2}{12}\left(\frac{1}{1+n_k}\right) \int
dP\,dR\ dP'\, (2\pi)^4
             \delta^4(P+P'-R-K) \nonumber \\
&&\rho_p\rho_r\rho_{p'}(1+n_p)(1+n_{p'})n_r\label{pi}
\eea

Rearranging we obtain \cite{meg},
\bea
&& I(k,k)_{lm} = \frac{\lambda^2}{12(1+n_k)} \int dP\,dR\,dP' ~(2\pi)^4
\delta({P} + {P'} - {R} - {K}) \rho_p \rho_r\rho_{p'}\label{Bint2} \\
&& ~~~~ [ B_{lm}({P}) + B_{lm}({P'}) - B_{lm}({K}) - B_{lm}({R})]
(1+n_p)(1+n_{p'})n_r
\nonumber
\eea
where we have used (\ref{def4}), (\ref{def5}) and (\ref{EQ1}).
When the delta functions are used to do the frequency integrals, this equation
has exactly the same form as the equation obtained from the
linearized Boltzmann equation (\ref{Bint}) with a shifted mass shell describing
effective thermal excitations.
Comparing (\ref{ttf11}) and (\ref{Bint}) with (\ref{ftf11}) and (\ref{Bint2})
we conclude that calculating  shear viscosity  using effective
 transport theory by keeping only first order terms in the gradient expansion
is equivalent to using the Kubo formula obtained from linear response theory,
with a three-point vertex obtained by resumming ladder graphs.  The
contributions to the viscosity are shown in Fig.~[3].
%%%%%%%%%%%%%%%%%%%%%%%%%%% FIG. 3 %%%%%%%%%%%%%%%%%%%%%%%%%%%%%%%%%%%%
%%%%%%%%%%%%%%%%%%
%%%%%%%%%%%%%%%%%%%%%%%%%%% %%%%%%%%%%%%%%%%%%%%%%%%%%%%%%%%%%%%%
\begin{eqnarray}
\parbox{14cm}
{{
\begin{center}
\parbox{10cm}
{
\epsfxsize=8cm
\epsfysize=3cm
\epsfbox{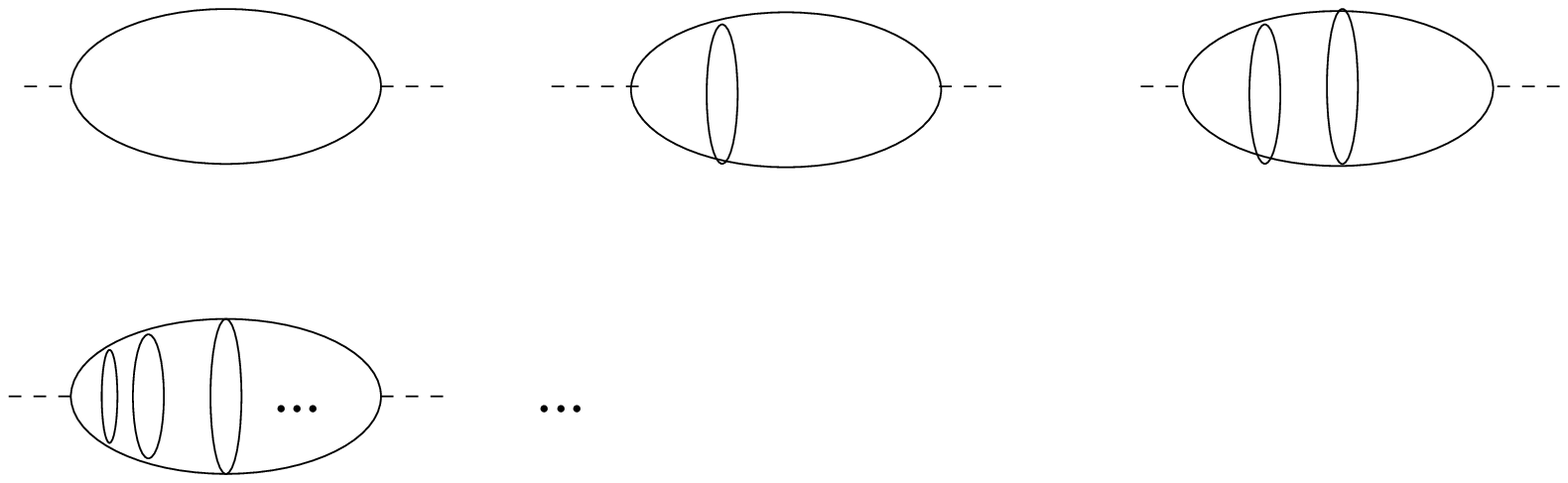}}\\
\parbox{14cm}{\small \center Fig.~[3]:  Some of the ladder diagrams that
contribute to shear viscosity.  }
\label{F3}
\end{center}
}}
\nonumber
\end{eqnarray}

\subsubsection{The Extended-Ladder Resummation}

In this section we consider an integral equation for the vertex $M_{ijlm}$
which resums an
infinite set of graphs that includes ladder graphs and some other contributions
which we will call extended ladder graphs.  We will show that this integral
equation has exactly the same form as the
integral equation (\ref{Cint}) obtained by expanding the Boltzmann equation to
second order.
We consider the integral equation shown in Fig.~[4].
%%%%%%%%%%%%%%%%%%%% FIG 4 %%%%%%%%%%%%%%%%%%%%%%%%%%%%%%%%%%%%%%%%%%%%%%%%%
 \begin{eqnarray}
\parbox{14cm}
{{
\begin{center}
\parbox{10cm}
{
\epsfxsize=8cm
\epsfysize=3cm
\epsfbox{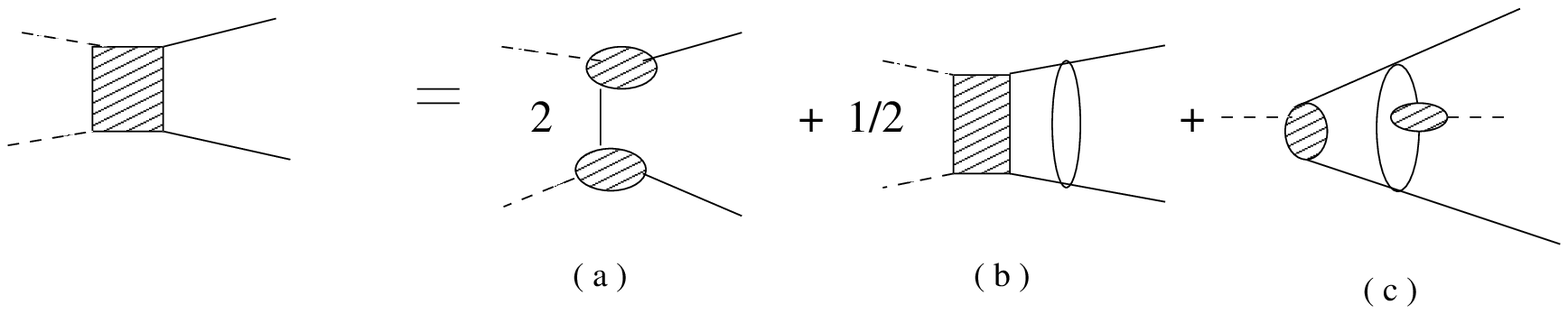}}\\
\parbox{14cm}{\small \center Fig.~[4]:  Integral equation for an
extended-ladder resummation.  }
\label{F4}
\end{center}
}}
\nonumber
\end{eqnarray}

Following (\ref{mfbar}) and (\ref{kubo2})
we calculate contributions to $M_F$, $M_{R1}$, $M_{R4}$ for each diagram. We
introduce some notation to simplify the equations.  For four-point vertices we
list the first three momenta only and for three-point vertices we list the
first two only.  In both cases the last momentum is the one that is determined
by the conservation of energy.  For example: $M(K,Q,Q',-K-Q-Q'):= M(K,Q,Q')$
and $\Gamma(K,Q,-K-Q):= \Gamma(K,Q)$. In addition we write (as before)
$M(K,0,0,-K):= M(K)$ and $\Gamma(K,0,-K):=\Gamma(K)$.
{}From Fig.~[4a] we have,
\bea
&&(M_F^{ijlm})^{4a} (K,Q,Q') = 2 i[ r_{k+q} \Gamma^{ij}_{R1} (K,Q)
\Gamma_{F2}^{lm}(K+Q,Q')
\nonumber \\
&&~~~~ +a_{k+q} \Gamma^{ij}_{F2} (K,Q) \Gamma_{R3}^{lm}(K+Q,Q')
+f_{k+q} \Gamma^{ij}_{R1} (K,Q) \Gamma_{R3}^{lm}(K+Q,Q') ]
\nonumber
 \\
&&(M_{R1}^{ijlm})^{4a} (K,Q,Q') = 2 i r_{k+q} \Gamma^{ij}_{R1} (K,Q)
\Gamma_{R1}^{lm}(K+Q,Q')
\nonumber
\\
&&(M_{R4}^{ijlm})^{4a} (K,Q,Q')= 2 i a_{k+q} \Gamma^{ij}_{R3} (K,Q)
\Gamma_{R3}^{lm}(K+Q,Q')  \label{ma}
\eea
%
%%%%%%%%%%%%%%%%%%%%%%%%%%%%%%%%%%%%%%%%%%%%%%%%%%%%%%%%%%%%
%
We can rewrite this result by substituting in the expanded form of the vertex
$\Gamma_{ij} = I_{ij} + \frac{1}{2} \tilde \Gamma_{ij}$ with 
$\tilde \Gamma_{ij}$  obtained from the integral equation that 
corresponds to Fig.~[2]. We use,
\bea
&&\Gamma_{R1}^{ij} (K,Q) \nonumber \\
&&~=I^{ij}(k) -
\frac{i\lambda^2}{8} \int dP \,dR\  r_p a_{p+q} (N'-N_r) a_r r'\,
  [\Gamma_{F2}^{ij}(P,Q)-N_{p+q}\Gamma_{R1}^{ij}(P,Q)
+N_p \Gamma_{R3}^{ij}(P,Q)]
\nonumber\\
&& \Gamma_{R3}^{ij} (K,Q) \nonumber \\
&&~ =I^{ij}(k) -
\frac{i\lambda^2}{8} \int dP \,dR\  r_p a_{p+q} (N_r-N') r_r a'\,
[\Gamma_{F2}^{ij}(P,Q)-N_{p+q}\Gamma_{R1}^{ij}(P,Q) +N_p \Gamma_{R3}^{ij}(P,Q)]
\nonumber
\\
&&\Gamma_{F2}^{ij} (K,Q) \nonumber \\
&&~ = - \frac{i\lambda^2}{8} \int dP \,dR\  r_p a_{p+q}
(1-N'N_r) \rho_r \rho' \,  [\Gamma_{F2}^{ij}(P,Q)-N_{p+q}\Gamma_{R1}^{ij}(P,Q)
+N_p \Gamma_{R3}^{ij}(P,Q)]
\nonumber
\eea
It is easy to show that the contribution from Fig.~[4a] is equivalent to the
contributions
 from the four diagrams in Fig.~[5].
%%%%%%%%%%%%%%%%%%%%%% FIG 5 %%%%%%%%%%%%%%%%%%%%%%%%%%%%%%%%%%%%%%%%%%%%%%%%%
 \begin{eqnarray}
\parbox{14cm}
{{
\begin{center}
\parbox{10cm}
{
\epsfxsize=8cm
\epsfysize=3cm
\epsfbox{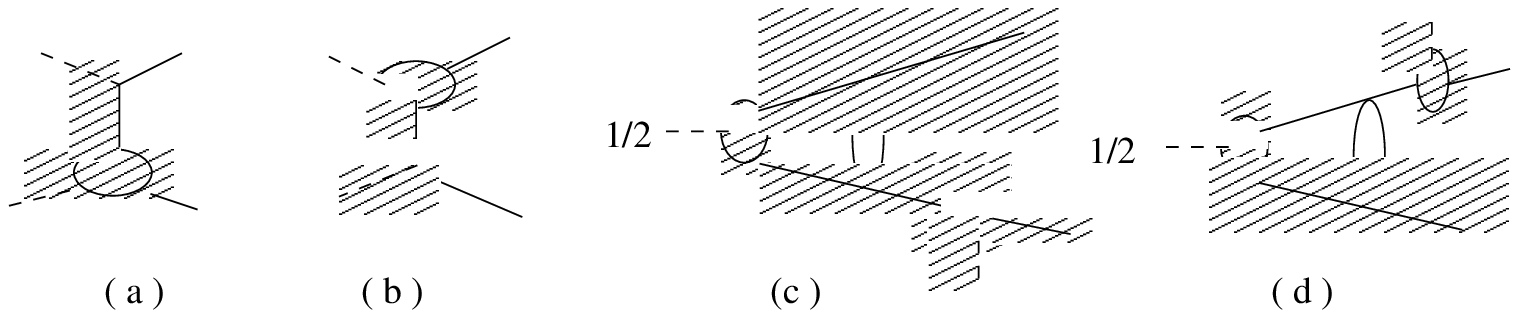}}\\
\parbox{14cm}{\small \center Fig.~[5]: The four-point vertices that correspond
to the diagram in  to Fig.~[4a]. }
\label{F5}
\end{center}
}}
\nonumber
\end{eqnarray}

\noindent From Figs.~[5a,5b] we obtain,
\bea
&&(M_F^{ijlm})^{5a} (K,Q,Q') =  iI_{ij}(k)[a_{k+q}\Gamma^{lm}_{F2}(K,Q)+
f_{k+q}\Gamma^{lm}_{R1}(K,Q)]
 \\
&&(M_{R1}^{ijlm})^{5a} (K,Q,Q') =  ir_{k+q}I_{ij}(k)\Gamma^{lm}_{R1}(K,Q)
 \\
&&(M_{R4}^{ijlm})^{5a} (K,Q,Q') =  ia_{k+q}I_{ij}(k)\Gamma^{lm}_{R3}(K,Q)
 \\
&&(M_F^{ijlm})^{5b} (K,Q,Q') =  iI_{ij}(k)[r_{k+q}\Gamma^{lm}_{F2}(K+Q,Q')+
f_{k+q}\Gamma^{lm}_{R3}(K,Q)]
 \\
&&(M_{R1}^{ijlm})^{5b} (K,Q,Q') =  ir_{k+q}I_{ij}(k)\Gamma^{lm}_{R1}(K+Q,Q')
 \\
&&(M_{R4}^{ijlm})^{5b} (K,Q,Q') =  ia_{k+q}I_{ij}(k)\Gamma^{lm}_{R3}(K+Q,Q')
\eea
{}From Figs.~[5c,5d] we obtain,
\bea
&&(M_F^{ijlm})^{5c}(K,Q,Q') \nonumber\\
&&~~~~=\frac{-i\lambda^2}{8}\int\,dP \,dR\,
a_{p+q+q'}r_p\nonumber \\
&&~~~~
[\Gamma^{lm}_{F2}(P+Q,Q')-N_{p+q+q'}\Gamma^{lm}_{R1}(P+Q,Q')
+N_{p+q}\Gamma^{lm}_{R3}(P+Q,Q')]\nonumber \\
&&~~~~\cdot [\Gamma^{ij}_{F2}(K,Q)(f_r a_{p'} +r_r
f_{p'})a_{k+q}\nonumber \\
&&~~~~ +\Gamma^{ij}_{R1}(K,Q) (f_{k+q}(f_r a_{p'} +r_r f_{p'})+r_{k+q}(f_{p'}
f_r +a_{p'} r_r+r_{p'} a_r))]\\
&& \nonumber\\
&&(M_{R1}^{ijlm})^{5c}(K,Q,Q')\nonumber\\
&&~~~~=\frac{-i\lambda^2}{8} \int\,dP \,dR\,a_{p+q+q'}r_p \nonumber \\
&&~~~~
[\Gamma^{lm}_{F2}(P+Q,Q')-N_{p+q+q'}\Gamma^{lm}_{R1}(P+Q,Q')+
N_{p+q}\Gamma^{lm}_{R3}(P+Q,Q')]\nonumber
\\
&&~~~~\cdot \Gamma_{R1}^{ij}(K,Q)r_{k+q}(f_{p'} a_r +r_{p'} f_r)
\\
&&\nonumber \\
&&(M_{R4}^{ijlm})^{5c}(K,Q,Q')\nonumber\\
&&~~~~=\frac{-i\lambda^2}{8}\int\,dP \,dR\, a_{p+q+q'}r_p \nonumber \\
&&~~~~
[\Gamma^{lm}_{F2}(P+Q,Q')-N_{p+q+q'}\Gamma^{lm}_{R1}(P+Q,Q')+
N_{p+q}\Gamma^{lm}_{R3}(P+Q,Q')]\nonumber
\\
&&~~~~
\cdot \Gamma^{ij}_{R3}(K,Q)a_{k+q}(f_{p'} r_r +a_{p'} f_r).
\\
\nonumber
\\
&&(M_F^{ijlm})^{5d}(K,Q,Q') \nonumber\\
&&~~~~= \frac{-i\lambda^2}{8}\int\,dP \,dR\,
a_{p+q} r_p [\Gamma^{lm}_{F2}(P,Q) - N_{p+q}\Gamma^{lm}_{R1}(P,Q) +
N_p\Gamma^{lm}_{R3}(P,Q)] \nonumber\\
&&~~~~~~[\Gamma^{ij}_{F2}(K+Q,Q') r_{k+q} (a_r f'+ f_r r') \nonumber \\
&&~~~~+ \Gamma^{ij}_{R3}(K+Q,Q')(f_{k+q}(a_r f' +f_r r') + a_{k+q}
(f_r f' + r_r a' + a_r r'))]  \\
&& \nonumber \\
&&(M_{R1}^{ijlm})^{5d}(K,Q,Q')\nonumber\\
&&~~~~ = \frac{-i\lambda^2}{8}\int\,dP \,dR\,
r_p a_{p+q} [\Gamma^{lm}_{F2}(P,Q) - N_{p+q}\Gamma^{lm}_{R1}(P,Q) + N_p
\Gamma^{lm}_{R3}(P,Q)] \nonumber \\
&&~~~~\cdot \Gamma^{ij}_{R1}(K+Q,Q') r_{k+q} (a_r f' + f_r r')  \\
&& \nonumber \\
&&(M_{R4}^{ijlm})^{5d}(K,Q,Q') \nonumber\\
&&~~~~= \frac{-i\lambda^2}{8}\int\,dP \,dR\,
r_p a_{p+q} [\Gamma^{lm}_{F2}(P,Q) - N_{p+q}\Gamma^{lm}_{R1}(P,Q) + N_p
\Gamma^{lm}_{R3}(P,Q)] \nonumber \\
&&~~~~\cdot \Gamma^{ij}_{R3}(K+Q,Q') a_{k+q}(a' f_r + f' r_r)
\eea
The contribution from Fig.~[4b] is
\bea
&&(M_F^{ijlm})^{4b}(K,Q,Q')=-\frac{\lambda^2}{8} \int\,dP \,dR\, a_{p+q+q'} 
r_p\bar M_F^{ijlm}(P,Q,Q')[r_r
a_{p'} +a_r r_{p'} +f_r f_{p'}]
 \\
&&(M_{R1}^{ijlm})^{4b}(K,Q,Q')=-\frac{\lambda^2}{8} \int\,dP \,dR\, a_{p+q+q'}
r_p \bar M_F^{ijlm}(P,Q,Q')[a_r f_{p'} +f_r
r_{p'}]\\
&&(M_{R4}^{ijlm})^{4b}(K,Q,Q')=-\frac{\lambda^2}{8} \int\,dP \,dR\, a_{p+q+q'}
r_p \bar M_F^{ijlm}(P,Q,Q')[r_r f_{p'} +f_r a_{p'}]
\eea
where $P'=R+K-P$.
Fig.~[4c] gives,
\bea
(M_F^{ijlm})^{4c}(K,Q,Q')=&&\frac{-i\lambda^2}{4} \int\,dP \,dR\, (N_p-N_{p+q})
(N_{r-q'}-N_r)
\\
\nonumber
&&
\Gamma_{R2}^{ij*}(P,Q)\Gamma_{R2}^{lm*}(R-Q',Q')
f_{p'} r_p a_{r+q} a_r r_{r-q'}\nonumber
\\
(M_{R1}^{ijlm})^{4c}(K,Q,Q')=&&\frac{-i\lambda^2}{4}\int\,dP \,dR\,
(N_p-N_{p+q}) (N_{r-q'}-N_r) \Gamma_{R2}^{ij*}(P,Q)
\\
\nonumber
&&\Gamma_{R2}^{lm*}(R-Q',Q') r_{r'} r_p a_{r+q}
a_r r_{r-q'}\nonumber
\\
(M_{R4}^{ijlm})^{4c}(K,Q,Q')=&&\frac{-i\lambda^2}{4} \int\,dP
\,dR\,(N_p-N_{p+q}) (N_{r-q'}-N_r) \Gamma_{R2}^{ij*}(P,Q)
\\
\nonumber
&&\Gamma_{R2}^{lm*}(R-Q',Q') a_{p'} r_p a_{r+q}
a_r r_{r-q'}
\eea

To combine these expressions we use (\ref{kms}) and (\ref{mfbar}). We keep only
the pinching contributions and use (\ref{pinch}) to regulate. As discussed
previously, we expand in $q_0$ and $q_0'$ and keep the term proportional to
$q_0q_0'$,  since that is the only term that will contribute to the quadratic
shear viscous response coefficient.  As before, one can show that the terms in
the expansion that contain gradients acting on the vertices $M_F,~M_{R1}$ and
$M_{R4}$ do not contribute to the result and that we can use (\ref{re2}).
We simplify further by using (\ref{nexp}).   We also make repeated use of the
set of identities below which hold for momenta which satisfy $P_1+P_2=P_3+P_4$:
\bea
&&n_1 n_2 (1+ n_3)(1+n_4) = (1+n_1)(1+n_2)n_3 n_4 \nonumber \\
&& n_3 n_{2}+ n_3 +n_1 (n_3-n_{2}) = (1+n_1)(1+n_{2})n_3 / (1+n_4) \\
&& n_1(1+n_1)n_3(1+n_3)(N_2-N_4) = (1+n_1)(1+n_2)n_3n_4(N_3-N_1)\,.\nonumber
\eea
Finally, since $K$ is the momentum for an external leg
we take the on shell piece:
\be
r_k  \rightarrow \frac{i}{{\rm Im} \Sigma_k}
\ee
We obtain:
\bea
&& N_k n_k(1+n_k) M_{R1}^{ijlm}(K) \nonumber \\
&&~~~~= - n_k(1+n_k)N_kI_{ij} \frac {\Gamma_{R2}^{lm}(K)}{{\rm Im}\Sigma_k}
+\frac{\lambda^2}{4}
\int \,dP\,dR\,(1+n_p)(1+n_{p'})n_r n_k\rho_p\rho_p'\rho_r
\nonumber\\
&& ~~~~ \left[
- \frac {N_{p} M_{R1}^{ijlm}(P)}{{\rm Im}\Sigma_{p}}
+\frac{1}{2} \frac {\Gamma_{R2}^{ij}(P)}{{\rm Im}\Sigma_{p}}\frac
{\Gamma_{R2}^{lm}(R)} {{\rm Im}\Sigma_{r}} (N_r-N_p)
 + \frac{1}{2}\frac {\Gamma_{R2}^{ij}(P)}{{\rm Im}\Sigma_{p}}\frac
{\Gamma_{R2}^{lm}(K)} {{\rm Im}\Sigma_k} (N_p-N_k)
\right]
\eea
We introduce the symmetric notation: $P:=P_1\,;~P':=P_2\,;~R:=P_3$ and rewrite
the equation above after symmetrizing on the integration variables.
Rearranging we obtain,
\bea
&& n_k(1+n_k)N_kI_{ij} \frac{\Gamma_{R2}^{lm}(K)}{{\rm Im}\Sigma_k} \nonumber
\\
&&~~~~=\frac{\lambda^2}{12}\int \int \,dP_1\,dP_2\,dP_3
{(2\pi)}^4\delta^4(P_1+P_2-P_3-K)(1+n_1)(1+n_2)n_3n_k \rho_1 \rho_2 \rho_3
\nonumber
\\
&&~~~~[\frac{N_{p_3} M_{R1}^{ijlm} (P_3) }{{\rm Im}\Sigma_{p_3}}
+\frac{N_{k} M_{R1}^{ijlm} (K) }{{\rm Im}\Sigma_k}
 - \frac{N_{p_1} M_{R1}^{ijlm} (P_1) }{{\rm Im}\Sigma_{p_1}}
- \frac{N_{p_2} M_{R1}^{ijlm} (P_2) }{{\rm Im}\Sigma_{p_2}}
\nonumber
\\
&&~~~~+\frac{1}{2}\{ (N_1+N_2)\frac{\Gamma_{R2}^{ij}(P_1)}{{\rm
Im}\Sigma_{p_1}}
\frac{\Gamma_{R2}^{lm}(P_2)}{{\rm Im}\Sigma_{p_2}}
-(N_k+N_3)\frac{\Gamma_{R2}^{ij}(K)}{{\rm Im}\Sigma_k}
\frac{\Gamma_{R2}^{lm}(P_3)}{{\rm Im}\Sigma_{p_3}}
\nonumber
\\
&&~~~~+(N_3-N_1)\frac{\Gamma_{R2}^{ij}(P_1)}{{\rm Im}\Sigma_{p_1}}
\frac{\Gamma_{R2}^{lm}(P_3)}{{\rm Im}\Sigma_{p_3}}
+(N_k-N_1)\frac{\Gamma_{R2}^{ij}(P_1)}{{\rm Im}\Sigma_{p_1}}
\frac{\Gamma_{R2}^{lm}(K)}{{\rm Im}\Sigma_k}
\nonumber
\\
&&~~~~+(N_3-N_2)\frac{\Gamma_{R2}^{ij}(P_3)}{{\rm Im}\Sigma_{p_3}}
\frac{\Gamma_{R2}^{lm}(P_2)}{{\rm Im}\Sigma_{p_2}}
+(N_k-N_2)\frac{\Gamma_{R2}^{ij}(K)}{{\rm Im}\Sigma_k}
\frac{\Gamma_{R2}^{lm}(P_2)}{{\rm Im}\Sigma_{p_2}}
    \}]
\label{ieq2}
\eea
Note that once again we have obtained an integral equation that is decoupled:
it only involves $M_{R1}$ and $\Gamma_{R2}$.  With $\Gamma_{R2}$
determined by the integral equation (\ref{Bint2}), Equation (\ref{ieq2})
can be solved to obtain $M_{R2}$.  From the diagrams in Figs.~[4,5] we see that
the solutions to the integral equation will contain contributions of the form
shown in Fig.~[6].  Finally, by using (\ref{def4}), (\ref{def5}), (\ref{EQ1})
and (\ref{EQ2}) and comparing (\ref{ftf}) and (\ref{ieq2}) with (\ref{ttf}) and
(\ref{Cint}) we see that calculating the quadratic shear viscous response using
 transport theory describing  effective thermal excitations and keeping terms
that are quadratic in the gradient of the four-velocity field in the expansion
of the Boltzmann equation, is equivalent to calculating the same response
coefficient from quantum field theory at finite temperature using the
next-to-linear response Kubo formula with a vertex given by a specific integral
equation. This integral equation shows that the complete set of diagrams that
need to be resummed includes the standard ladder graphs, and an additional set
of extended ladder graphs. Some of the diagrams that contribute to the
viscosity are shown in Fig.~[7].

\begin{eqnarray}
\parbox{14cm}
{{
\begin{center}
\parbox{10cm}
{
\epsfxsize=8cm
\epsfysize=6cm
\epsfbox{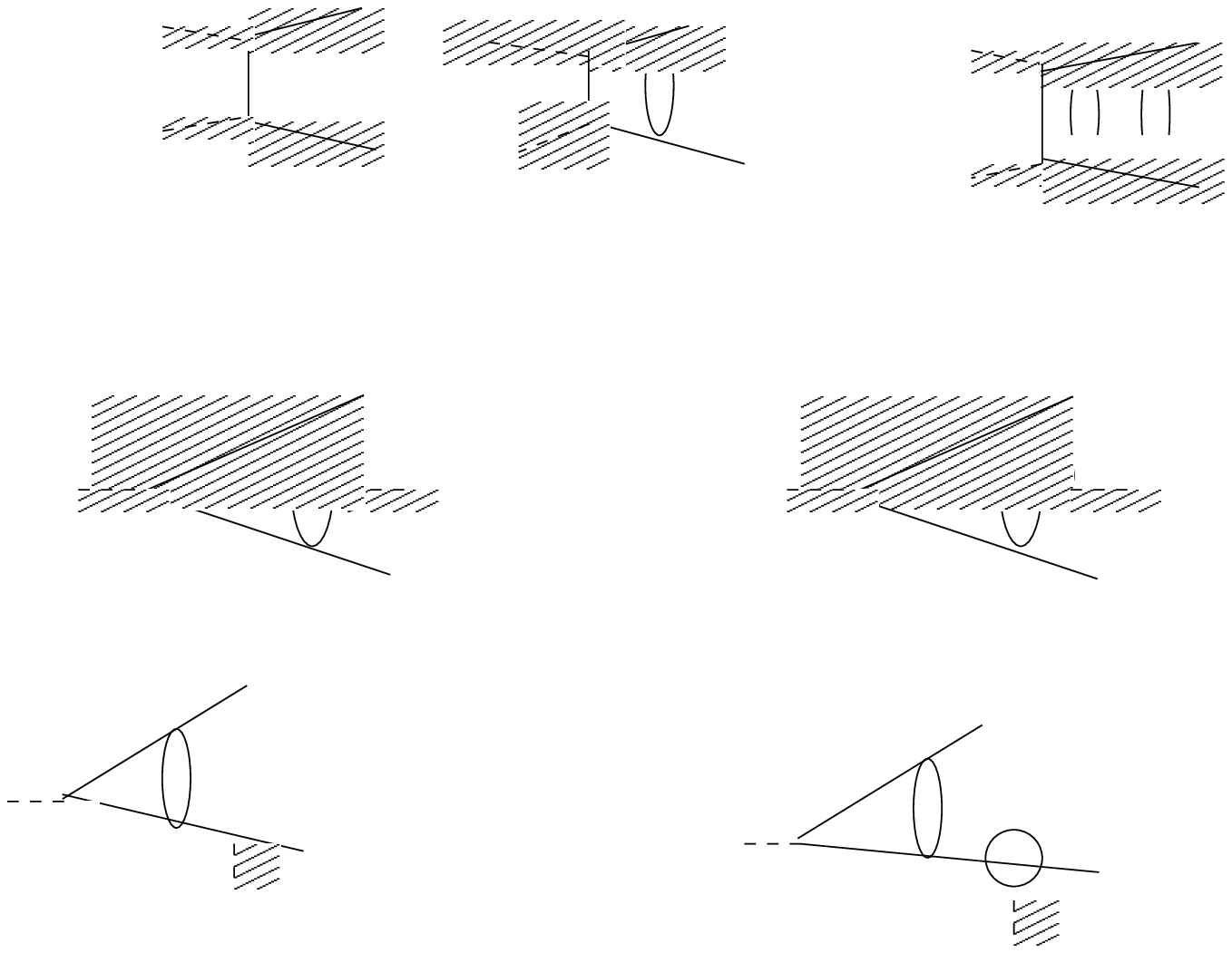}}\\
\parbox{14cm}{\small \center Fig.~[6]:  Some of the ladder and extended-ladder
diagrams that contribute to the four-point vertex. }
\label{F6}
\end{center}
}}
\nonumber
\end{eqnarray}
\begin{eqnarray}
\parbox{14cm}
{{
\begin{center}
\parbox{10cm}
{
\epsfxsize=8cm
\epsfysize=4cm
\epsfbox{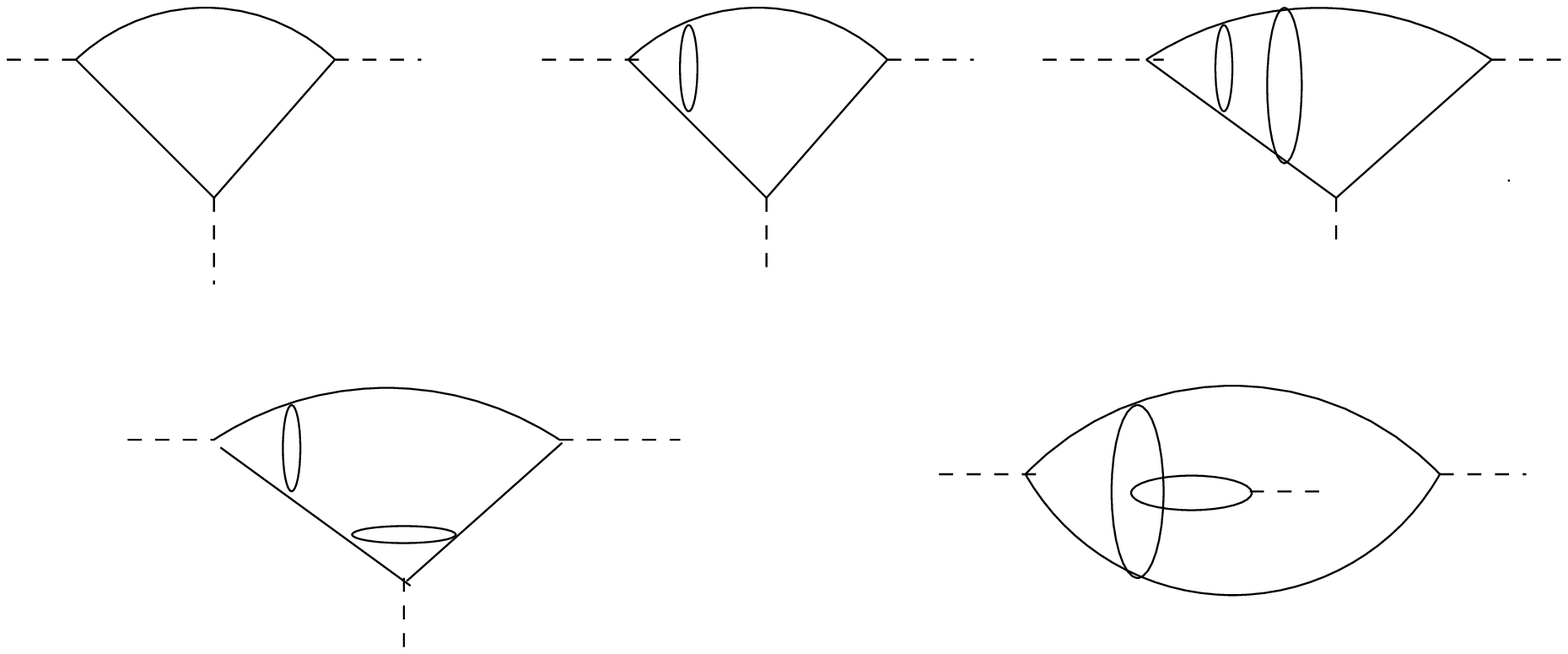}}\\
\parbox{14cm}{\small \center Fig.~[7]:  Some of the ladder and extended ladder
diagrams that contribute to quadratic shear viscous response. }
\label{F7}
\end{center}
}}
\nonumber
\end{eqnarray}

\section{Conclusions}
\label{V}
We have studied nonlinear response using two different methods.  The first
method uses standard transport theory.  We start from a local equilibrium form
for the distribution function and perform a gradient expansion.  We calculate
the quadratic shear viscous response  coefficient by expanding the Boltzmann
equation and obtaining a hierarchy of equations that can be solved
consistently.
  The second technique uses response theory.  We work with a perturbative
Hamiltonian that is linear in the gradient of the four-velocity field and study
quadratic response. We generalize the Kubo formula for linear response
and obtain an expression that allows us to calculate quadratic shear viscous
response from the  retarded three-point green function of the viscous shear
stress tensor. The transport theory calculation  involves the use of
the Boltzmann equation which is itself obtained from some more fundamental
theory. The response theory calculation uses the well known
methods of equilibrium finite temperature quantum field theory and is, in this
sense, more fundamental.  However, the response theory calculation is
complicated by the need to resum infinite sets of diagrams at finite
temperature.

At leading order, it is well known that a correct calculation of the linear
response coefficient involves the resummation of ladder graphs.  Beyond leading
order in response theory it is difficult even to identify which diagrams need
to be resummed.
We have identified precisely which diagrams need to be
resummed by studying the connection between the transport theory calculation
and the response theory calculation. We have shown that calculating the
quadratic  shear viscous response coefficient using transport theory by keeping
terms that are quadratic in the gradient of the four-velocity field in the
expansion of the Boltzmann equation, is equivalent to calculating the same
response coefficient from quantum field theory at finite temperature using the
next-to-linear response Kubo formula with a vertex given by a specific integral
equation. This integral equation shows that the complete set of diagrams that
need to be resummed includes the standard ladder graphs, and an additional set
of extended ladder graphs.

There are several directions for future work.  It has  been shown that the
Boltzmann equation can be derived from the Kadanoff-Baym equations by using a
gradient expansion and keeping only linear terms \cite{SL}.  The connection
between this result and the work discussed in this paper can probably be
understood by studying the dual roles of the gradient expansion and the
quasiparticle approximation.  In addition, it would be interesting to
generalize this work to gauge field theories.

%%%%%%%%%%%%%%%%%%%%%%%%%%%%%%%%%%%%%%%%%%%%%%%%%%%%%%%%%%%%
\acknowledgments
%%%%%%%%%%%%%%%%%%%%%%%%%%%%%%%%%%%%%%%%%%%%%%%%%%%%%%%%%%%%

 This work was supported by the Natural Sciences and Engineering 
and Research Council of Canada (NSERC), and the National Natural Science 
Foundation of China (NSFC).

\end{document}